\documentclass{article}
\usepackage{spconf,amsmath,graphicx}

\usepackage[T1]{fontenc} 
\usepackage[utf8]{inputenc} 
\usepackage{amsmath,cite,url, amssymb}
\usepackage{pifont}
\newcommand{\cmark}{\textcolor{green}{\ding{51}}}%
\newcommand{\xmark}{\textcolor{red}{\ding{55}}}%
\usepackage{graphicx}
\usepackage{color}
\usepackage{float}
\usepackage{multirow}

\title{PERFORMANCE CONDITIONING FOR DIFFUSION-BASED MULTI-INSTRUMENT MUSIC SYNTHESIS}
\name{Ben Maman$^{\star}$ \qquad Johannes Zeitler$^{\dagger}$ \qquad Meinard M{\"u}ller$^{\dagger}$ \qquad Amit H. Bermano$^{\star}$\thanks{The International Audio Laboratories Erlangen are a joint institution of the Friedrich-Alexander-Universit{\"a}t Erlangen-N{\"u}rnberg (FAU) and the Fraunhofer Institute for Integrated Circuits IIS.}}
\address{$^{\star}$ Tel Aviv University \\
  $^{\dagger}$International Audio Laboratories Erlangen, Germany}

\ninept
\begin{document}
%
\maketitle
\begin{abstract}
Generating multi-instrument music from symbolic music representations is an important task in Music Information Retrieval (MIR). A central but still largely unsolved problem in this context is musically and acoustically informed control in the generation process. As the main contribution of this work, we propose enhancing control of multi-instrument synthesis by conditioning a generative model on a specific performance and recording environment, thus allowing for better guidance of timbre and style. Building on state-of-the-art diffusion-based music generative models, we introduce \textit{performance conditioning} -- a simple tool indicating the generative model to synthesize music with style and timbre of specific instruments taken from specific performances. Our prototype is evaluated using uncurated performances with diverse instrumentation and achieves state-of-the-art FAD realism scores while allowing novel timbre and style control. Our project page, including samples and demonstrations, is available at \url{benadar293.github.io/midipm}.

\end{abstract}


\section{Introduction}\label{sec:introduction}
Multi-Instrument Music Synthesis is the task of generating audio from MIDI files, emulating specific instruments played with desired notes and timbre. It is a novel task in Music Information Retrieval (MIR), attracting increasing attention in recent years, with applications in music creation and production for all proficiency levels. 
It comprises several challenges, one of the central thereof is control over the style of the synthesized pieces.

Since physically modeling and specifying the different sound phenomena (e.g., vibrato, intensity, echo, and specific instrument timbre) for generation is unfeasible, current approaches avoid the flat and unrealistic sound of traditional concatenative synthesizers and naturally infer the aspects of generation from data, 
typically using denoising diffusion probabilistic models (DDPMs) 
~\cite{DBLP:conf/ismir/HawthorneSRZGME22, DBLP:journals/corr/abs-2302-03917}.
However, the subtle nuances of expression are often lost in the generation process, resulting in less realistic audio, and other phenomena such as instrument drift, where the same instrument is not rendered coherently in different parts of the generated piece.

In this work, we introduce a mechanism that specifically tackles style control; 
Except for conditioning the model on notes and instruments, our synthesizer is also conditioned on the performance, enabling generation with performance-specific characteristics, such as timbre, style, and 
recording environment. This \textit{performance conditioning} not only increases realism but also allows the sound of, e.g., a specific guitar to be reproduced in a specific acoustic environment. 

To give a concrete example, our approach enables reproducing the sound of 
the guitar in a 1975 recording of Segovia playing Alb{\'e}niz's Capriccio Catal{\'a}n, now playing another piece, such as Jobim's Felicidad. 
To the best of our knowledge, our work is the first to offer this capability in a multi-instrument setting.

Performance conditioning is integrated, like the iteration parameter $t$ of the diffusion process, using FiLM layers~\cite{DBLP:conf/aaai/PerezSVDC18}, and a sampling-overlapping technique ensures smooth transitions between generated segments, together providing coherent generation of long sequences. 





Through extensive evaluation with established score metrics~\cite{DBLP:conf/interspeech/KilgourZRS19}, e.g., Fr\'{e}chet Audio Distance (FAD), we show that our concept enhances perceptual similarity to the desired performance while improving realism. 

\section{Related Work}
Audio synthesis in current literature can be done auto-regressively, where models directly construct a waveform sample-by-sample~\cite{DBLP:conf/ssw/OordDZSVGKSK16, DBLP:conf/icml/EngelRRDNES17,DBLP:conf/iclr/HawthorneSRSHDE19}. 
Another approach, which we take, operates in the spectral domain. This requires a subsequent step to convert the generated spectral representation (STFT, mel) into a waveform, but it is computationally more efficient.
For a data-driven approach, large amounts of labeled data are required, i.e., paired datasets of audio recordings and their corresponding time-aligned transcriptions.
While such datasets exist for instruments such as the piano thanks to special equipment (e.g., Disklavier), this is not the case for other instruments.
Thus, previous works mainly focus on the generation of piano performances, monophonic (single-voice) music,
or music produced by a concatenative synthesizer~\cite{DBLP:conf/ismir/HawthorneSRZGME22}. 

Table~\ref{table:compare} provides a summary of existing methods.~\cite{DBLP:conf/aaai/WangY19} uses a U-Net to synthesize solo violin, cello, and flute performances, requiring a separate model for each instrument. 
~\cite{DBLP:conf/icassp/DongZBM22} uses a Transformer architecture to synthesize solo violin or piano. Both works produce only monophonic and single-instrument music (i.e., only one note played by a specific instrument is synthesized at any given time).

~\cite{DBLP:conf/iclr/WuMDSKCCHE22} 
learns a parametric model of a musical performance, 
synthesizing from controls such as intensity, vibrato, etc. Although promising, the main drawback is the requirement of elaborate performance controls at inference time (vibrato etc.), which requires much effort even from a skilled musician. Instead, we rely on the diffusion model to generate such aspects of expression. In addition, ~\cite{DBLP:conf/iclr/WuMDSKCCHE22}
only operate on monophonic single-instrument data, similar to former works, due to a lack of high-quality polyphonic training data. 


In multi-instrument synthesis, Hawthorne et al.~\cite{DBLP:conf/ismir/HawthorneSRZGME22} use a T5 Transformer-based diffusion model. 
While this method is promising and produces high-fidelity audio, it has several limitations: It does not have control over the performance characteristics and style (e.g., type of organ, recording environment) and produces less realistic sound, 
as shown on our supplemental website with audio examples at
\url{benadar293.github.io/midipm}.

\section{Method}
\begin{figure*}
\centering
\includegraphics[width=0.9\linewidth]{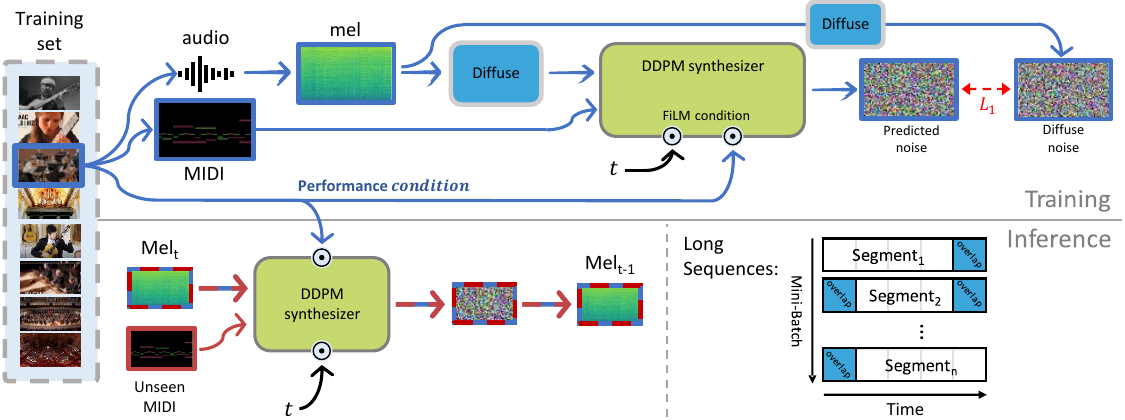}
\caption{
Overview of our proposed diffusion-based synthesis model. Performance conditioning (determining the style, recording environment, and specific timbre) is done through FiLM layers at each block, which can be applied to both to a T5 transformer and a U-Net. The performance condition ID is inserted at each layer by concatenation with the diffusion timestep.
}
\label{fig:overview}
\end{figure*}

\begin{table}
\begin{center}
\begin{tabular}{|c|c|c|c|c|c|}
\hline
 & \textbf{Multi} & \textbf{Perf.} & \textbf{Symph.} & \textbf{Data} & \textbf{Real\%} \\
\hline
\cite{DBLP:conf/iclr/HawthorneSRSHDE19} & \xmark & \cmark & \xmark & $\sim$140H & 100\% \\
\hline
~\cite{DBLP:conf/aaai/WangY19} & \xmark & \xmark & \xmark & $\sim$1H & 100\% \\
\hline
\cite{DBLP:conf/icassp/KimBKB19} & \xmark & \xmark & \xmark & $\sim$1H & 0\% \\
\hline
~\cite{DBLP:conf/icassp/DongZBM22} & \xmark & \xmark & \xmark & $\sim$1H & 100\% \\
\hline
~\cite{DBLP:conf/iclr/WuMDSKCCHE22} & \xmark & \xmark & \xmark & $\sim$3H & 100\% \\
\hline
\cite{DBLP:conf/ismir/HawthorneSRZGME22} & \cmark & \xmark & \xmark & $\sim$1500H & $\sim 2\%$ \\
\hline
Ours & \cmark & \cmark & \cmark & $\sim$58H & 100\% \\
\hline
\end{tabular}
\end{center}
\caption{Overview of previous work indicating the ability to render multiple instruments simultaneously (\textbf{Multi}), reproducing specific performance style (\textbf{Perf.}), generating orchestral symphonies \textbf{Symph.}), \textbf{Data} size, and ratio of real vs. synthetic data used for training (\textbf{Real\%}).
}
\label{table:compare}
\end{table}

An overview of our method is depicted in Figure~\ref{fig:overview}. We seek to enhance the generation quality and control of an off-the-shelf diffusion-based music generator using performance conditioning. Hence, starting from a dataset $\mathcal{D}=\{(a_i, m_i, p_i)\}_{i=1}^N$, comprising audio performances $a_i$, their symbolic MIDI annotation $m_i$, and information regarding the identity of the performing ensemble and recording environment $p_i$, we train a music synthesizer using state-of-the-art architectures  (see Section~\ref{subsection:architecture}), infused with performance conditioning. 

As previously mentioned, we choose to operate in the spectral domain, using the mel-spectogram representation, mainly for computational purposes. We postulate our method can be adapted to larger spectral representations (e.g., STFT), or the waveform domain, at significantly higher computational costs. To convert mel-spectrograms into audio, we rely on the state-of-the-art GAN-based Soundstream vocoder~\cite{zeghidour2021soundstream} (which is the same vocoder used by Hawthorne et al.~\cite{DBLP:conf/ismir/HawthorneSRZGME22}). We represent a performance condition ID $p_i$ as an integer number, where recordings performed by the same ensemble in the same recording environment are assigned the same number. Each condition can represent a single recording of a few minutes (e.g., Segovia
playing Albéniz’s Capriccio Catalán on the guitar) in the training set, or a set of
recordings of several hours (e.g., of Beethoven’s concertos for piano and orchestra performed by Mitsuko Uchida and The Orchestra of The Bavarian Radio). Performance conditioning is implemented using FiLM layers~\cite{DBLP:conf/aaai/PerezSVDC18}, a popular technique used to condition the denoising diffusion process on the iteration parameter $t$ (see Section~\ref{sec:conditioning} for more details).

During training, the synthesizer is trained to reconstruct performances, based on their MIDI representation and corresponding performance condition ID. During inference, new unseen MIDI performances are combined with different performance conditions that correspond to the performances seen during training to yield note control from the MIDI representation, and style and timbre from the performance conditioning.

Finally, to generate longer sequences and ensure their seamless transitions between generated intervals,
we adapt an overlapping technique, borrowed from visual generation (Section~\ref{sec:temporal}). 

\subsection{Architecture}\label{subsection:architecture}
We experiment with two architectures: A U-Net originally used for images~\cite{NEURIPS2020_4c5bcfec}, and a T5 Transformer~\cite{DBLP:conf/ismir/HawthorneSRZGME22}.

We adapt the \textbf{U-Net} to spectrogram synthesis by using 1D convolution, attention, and group normalization
rather than 2D, regarding the frequencies as channels. This is more expressive than spatial 2D convolutions,
as it captures interactions between distant frequencies, inherent in spectrograms (partial frequencies). It is a common practice in spectrogram synthesis (e.g.~\cite{DBLP:conf/aaai/WangY19} use a 1D U-Net without diffusion).

The \textbf{T5}, borrowed from Hawthorne et al.~\cite{DBLP:conf/ismir/HawthorneSRZGME22}, comprises a transformer encoder, encoding the MIDI condition, and a transformer decoder, processing the noise itself. It receives the encoded MIDI through cross-attention layers.

As can be seen in Section~\ref{section:experiments}, the SOTA transformer-based architecture indeed achieves slightly better results, but the U-Net architecture requires significantly less training time.

\subsection{Conditioning}
\label{sec:conditioning}

For \textbf{Performance Conditioning}, 
we apply FiLM layers. FiLM layers~\cite{DBLP:conf/aaai/PerezSVDC18} apply learned conditional affine transformations on network features. They are suitable for multi-modal cases, where a model learns many similar tasks simultaneously. They enable the different tasks to share most parameters while maintaining flexibility.

We apply them by predicting an affine transformation for each block of the network (T5 or U-Net), using MLPs. For the U-Net and T5 decoder, we concatenate the diffusion timestep representation with the performance ID representation.
We observed that conditioning also the T5 note encoder on performance using FiLM layers produced better results than conditioning the T5 decoder alone. 

For \textbf{Note Conditioning},
we insert the conditioning notes into the encoder in the T5 model, and into the input layer in the U-Net. The note input contains the information of note \textit{onset} (beginning time), \textit{instrument}, and approximate \textit{duration}.


\subsection{Temporal Coherency \& Smooth Transitions}\label{sec:temporal}
We generate long performances of several minutes, by segments of $\sim$5 seconds each (dictated by memory constraints).
For smooth transition between segments, we generate the segments with short overlaps, smoothly interpolating between consecutive segments, in each step of the sampling process. We use an interpolation coefficient linearly decreasing from $1$ to $0$ along the overlap, in the predicted sample $x_0 = \frac{x_t - \sigma_t\epsilon}{\alpha_t}$ of each step, derived from the predicted noise $\epsilon$. 
Borrowed from motion generation~\cite{tseng2022edge,shafir2023human}, this is an effective and convenient approach, performed solely at the sampling stage, and requiring no additional training components, contrary to~\cite{DBLP:conf/ismir/HawthorneSRZGME22}.
\subsection{Diffusion Process}
We use a cosine noise schedule,
with $T=1000$ training steps, and $T=250$ sampling steps. We use \textbf{Classifier-Free Guidance (CFG)} to control the conditioning strength, for both performance and notes -- we train with condition dropout of probability $0.1$, for notes and performance independently, and sample with guidance weights of $1.25$ for both, which gave best overall results after a parameter search.

\section{Experiments and Evaluation}\label{section:experiments}
In the following section, we first give an overview of the datasets used. Next, we present experimental results showing the potential of our proposed method in correctly synthesizing MIDI files, as well as the effect of performance conditioning on reproducing a desired timbre and style. Lastly, we briefly describe the listening examples which can be found on our supplemental website.
\subsection{Datasets}
We train on 197 performances of western classical music, comprising 19 instruments (including symphonies, chamber music, solo and other instrumentations), totaling in 58:06:07 hours. The data consists of performances from YouTube~\cite{youTube} and Musopen~\cite{Museopen}, with corresponding MIDI transcriptions from \url{www.kunstderfuge.com}, aligned as proposed by Maman and Bermano~\cite{DBLP:conf/icml/MamanB22}. Following~\cite{DBLP:conf/icml/MamanB22}, we augment the data by pitch-shifting up to $\pm2$ semitones. 
We label the data with performance IDs by assigning
numerical indices to the different performances, where typically the same index is given to an entire set of recordings (e.g., a CD box with Beethoven's Piano Trios recorded by the same ensemble in the same studio).

We evaluate our models with 58 MIDI performances of western classical pieces of a total duration of 5:09:30 hours, none of which appear in the train set, but containing the same instruments.
For each test MIDI, we
randomly sample 3 conditioning performances
for synthesis.
For example, the test MIDI can be of Mozart's 40th symphony and the condition 
can be the performance of the Berlin Philharmonic Orchestra playing Brahms' Haydn Variations. See our website for the complete ensemble distribution of the data.
\subsection{Fr\'{e}chet Audio Distance (FAD)}
To evaluate the fidelity and resemblance of our generated performances to the conditioning performances, we use the \textbf{Fr\'{e}chet Audio Distance (FAD)}~\cite{DBLP:conf/interspeech/KilgourZRS19} -- a perceptual score with origins in computer vision~\cite{DBLP:conf/ismir/DefferrardBVB17}. FAD is based on large DNN-based models such as TRILL~\cite{DBLP:conf/interspeech/ShorJMLTQTSEH20}, trained on large real-world datasets to predict embedding vectors from snippets of input audio. 
The assumption is that perceptually similar audio snippets yield closely spaced embedding vectors. To compute FAD between two datasets (e.g., a set $\mathcal{D}_1$
of MIDI files synthesized
with a specific performance condition, and a set $\mathcal{D}_2$ of real recordings of the same performance), the mean vectors $\mu_{1,2}$ and the covariance matrices $\Sigma_{1,2}$ are computed over all embedding vectors generated from the respective datasets. The FAD is then defined as:
\begin{equation*}
    \mathrm{FAD}(\mathcal{D}_1, \mathcal{D}_2) = \lvert \mu_1 - \mu_2\rvert^2 + \mathrm{tr}\left( \Sigma_1 +\Sigma_2 - 2(\Sigma_1\Sigma_2)^{1/2} \right)\,.
\end{equation*}
Kilgour et al.~\cite{DBLP:conf/interspeech/KilgourZRS19} show that FAD correlates with human perception and that increasing distortions increase the FAD.
We use two models as backbones for FAD, also used by Hawthorne et al.~\cite{DBLP:conf/ismir/HawthorneSRZGME22}: TRILL~\cite{DBLP:conf/interspeech/ShorJMLTQTSEH20} (5.9 embeddings/sec.), and VGGish~\cite{DBLP:conf/icassp/HersheyCEGJMPPS17} (1 embedding/sec.). We measure FAD in two ways, differing in the choice of the compared datasets: \textbf{All}
(Section~\ref{subsection:all-fad}),
and \textbf{Group}
(Sect.~\ref{subsection:group-fad}).

\subsubsection{All-FAD - General Quality}
\label{subsection:all-fad}
To measure quality and fidelity, we measure FAD comparing the entire evaluation set to the entire train set. This measures the general similarity of the synthesized performances to real performances, rather than resemblance to a specific performance. We refer to this metric as \textbf{All-FAD} (Table~\ref{table:fad_transcription}). While this metric is important to measure fidelity, the main metric for evaluating performance conditioning is the \textbf{Group-FAD}
(Section~\ref{subsection:group-fad}, Table~\ref{table:fad_group}).

\subsubsection{Group-FAD - Resemblance to Target Performance}
\label{subsection:group-fad}
\begin{figure}[t!]
\centering
\includegraphics[width=0.99\columnwidth]{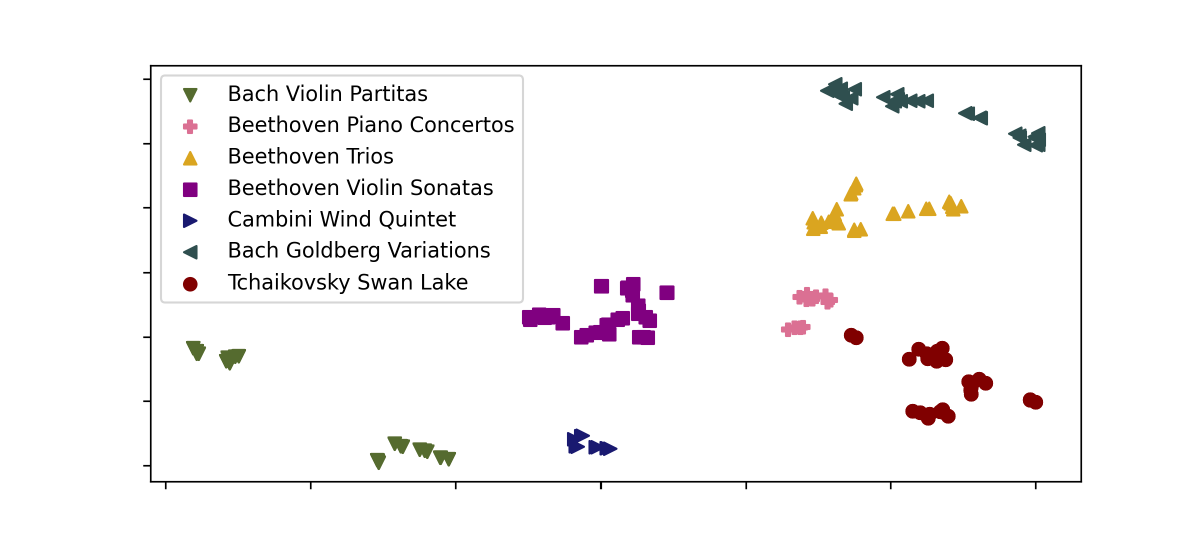}
\caption{
T-SNE Visualization of the TRILL embedding space. Points represent audio tracks, and colors represent complete recordings of specific performance IDs, comprising multiple such tracks. 
}
\label{fig:TSNE}
\end{figure}
Figure~\ref{fig:TSNE} shows a t-SNE visualization from the train set's TRILL embedding distribution. Each point represents the mean embedding of an audio track (e.g., a movement in a symphony), and each color represents a recording, comprising multiple such tracks. It
can be seen that tracks of the same performance form close clusters.

Following this insight, we define the \textbf{Group-FAD} metric (Table~\ref{table:fad_group}): To measure how well our synthesized performances resemble the target conditioning performance in timbre, room acoustics, etc., we compute FAD comparing each performance synthesized with a performance condition $p$, to the subset of the train set corresponding to $p$.
Furthermore, we use this metric to classify our generated performances, according to Group-FAD-nearest over all training performances.
\subsection{Correct Reproduction of Notes and Instruments}
To evaluate whether our models render the MIDI files correctly, i.e., whether the generated audio files contain the right notes at the right time, played by the correct instruments, we use \textbf{transcription} metrics: We measure the transcription accuracy of the synthesized performances using a transcriber (as in~\cite{DBLP:conf/iclr/HawthorneSRSHDE19, DBLP:conf/icml/MamanB22}) trained on the same data as the synthesizer. We compare the note events specified by the input MIDI to the transcription of the synthesized performance and measure the F1 score for \textbf{note} (pitch + onset within 50ms) and \textbf{note-with-instrument} (also correct instrument). 
\subsection{Results}
Results are reported in Tables~\ref{table:fad_transcription} (All-FAD and transcription), and~\ref{table:fad_group} (Group-FAD). Note that for FAD, the smaller the value, the greater the similarity (as desired). For the T5 model, for example, in Table~\ref{table:fad_transcription}, the TRILL-based All-FAD is 0.12 without performance conditioning and improves to 0.09 when performance conditioning is used. Similar tendencies (with the exception of U-Net and VGGish) can also be observed for the other models and the VGGish-based FAD. 

Next, we look at the transcription accuracy, indicating whether the synthesized performances actually realize the notes specified by the input MIDI files. As shown in Table 2, the T5 model reaches an accuracy of 63$\%$ (note-level), which is of reasonable magnitude when considering the complexity of highly polyphonic orchestral music. This is slightly higher than w/o performance conditioning, however, the more significant impact of performance conditioning on transcription is on the note-with-instrument level, which reaches 47$\%$ (w/) compared to 36$\%$ (w/o), indicating instrument identity is better preserved with performance conditioning. For the U-Net, performance conditioning does not significantly impact transcription.

With the All-FAD and transcription metrics, we discussed the quality of our model's generated performances, in terms of general similarity to real performances, and producing the desired notes. We now discuss Group-FAD, to evaluate performance conditioning, which is the main focus of our work. 
It can be seen in 
Table~\ref{table:fad_group} that performance conditioning consistently improves Group-FAD. For example, the VGGish-based Group-FAD drops from 7.1 (w/o) to 5.1 (w/) for the U-Net.

To get another perspective on the potential of our conditioning strategy, we performed a classification experiment. For each test performance synthesized with a conditioning performance $p$, we search for the performance in the train set, that is its nearest neighbor in terms of Group-FAD. 
As shown in Table 3, the top-1 classification accuracy is only 36$\%$ for the model T5 without conditioning. When using conditioning, it increases dramatically to 68$\%$. Similar improvements can also observed for the U-Net, and when looking at top-3 accuracy values. 
These results suggest that conditioning helps adapt to the specific timbre and room acoustics of a performance.

When comparing the actual FAD values in Tables~\ref{table:fad_transcription},~\ref{table:fad_group}, one can see the All-FAD values are lower than the Group-FAD. We attribute this to the fact that the 
mean vectors and covariance matrices for All-FAD are computed over significantly larger evaluation and reference datasets than for Group-FAD and therefore yield less statistical fluctuations, resulting in an overall lower FAD score.
\begin{table}[t!]
\begin{center}
\begin{tabular}{|c|c|c|c|c|c|c|c|c|}
\hline
\multirow{2}{*}{} & \multicolumn{4}{|c|}{All-FAD$\downarrow$} & \multicolumn{2}{|c|}{Transcription}\\
\cline{2-5}
& \multicolumn{2}{|c|}{VGGish} & \multicolumn{2}{|c|}{TRILL} & \multicolumn{2}{|c|}{N/N+I$\uparrow$} \\
\hline
P Con. & w/o & w/ & w/o & w/ & w/o & w/ \\
\hline
T5 & 3.9 & \underline{3.5} & 0.12 & \textbf{0.09} & 62/38\% & \textbf{\underline{63}}/\textbf{\underline{47}}\% \\
U-Net & \textbf{3.4} & 3.9 & 0.12 & \underline{0.11} & \textbf{\underline{63}}/\textbf{\underline{47}}\% & 62/46\% \\ 
\hline
\end{tabular}
\end{center}
\caption{Results for All-FAD, and transcription accuracy for note (N) and note-with-instrument (N+I). For each metric, the best result is bold, and the next-best is underlined.
}
\label{table:fad_transcription}
\end{table}
\begin{table}[t!]
\begin{center}
\begin{tabular}{|c|c|c|c|c|c|c|}
\hline
& \multicolumn{4}{|c|}{Group-FAD$\downarrow$} & \multicolumn{2}{|c|}{Perf. Acc.\%} \\
\cline{2-5}
 & \multicolumn{2}{|c|}{VGGish} & \multicolumn{2}{|c|}{TRILL} & \multicolumn{2}{|c|}{Top-1/3$\uparrow$} \\
\hline
P Con. & w/o & w/ & w/o & w/ & w/o & w/ \\
\hline
T5 & 5.8 & \underline{5.5} & 0.43 & \underline{0.35} & 36/60\% & \textbf{68}/\textbf{90}\% \\
U-Net & 7.1 & \textbf{5.1} & 0.5 & \textbf{0.33} & 14/30\% & \underline{56}/\underline{73}\% \\

\hline
\end{tabular}
\end{center}
\caption{Results for Group-FAD, and performance classification accuracy with TRILL FAD-nearest (Acc.\%). The best result in each metric is bold, and the next-best is underlined.
}
\label{table:fad_group}
\end{table}

While the quantitative analysis in this section indicates that performance conditioning indeed improves perceptual similarity in music synthesis, it cannot replace listening to the actual generated samples. Therefore, in order to assess the potential of performance conditioning in the context of diffusion-based music synthesis, we strongly encourage the reader to listen to the samples provided on our supplemental website
(\url{benadar293.github.io/midipm}).
We provide comparisons of MIDI files sonified with a simple concatenative synthesizer, the baseline approach~\cite{DBLP:conf/ismir/HawthorneSRZGME22}, and our proposed method. Furthermore, we show the conditioning effect when synthesizing the same MIDI file with a variety of different performance conditions. Among others, we render Bach's
8th Invention on eight different harpsichords, and Beethoven's
Pastoral Symphony with four different orchestras and recording
environments.
\section{Discussion}
We presented a framework for training neural synthesizers on real performances,
using diffusion models conditioned on notes and a performing style. We demonstrated that the latter condition both improves realism of multi-instrument performances of classical music, including orchestral symphonies, and adapts to the specific characteristics of a given performance, such as timbre and recording environment. Important 
future work includes extension to other genres, such as jazz, ethnic, pop music, and even human singing. Another important direction is exploring other spectral domains, such as STFT, CQT, etc. 
Yet another direction involves human speech -- 
we believe a unified diffusion-based framework for music and speech is possible, by providing additional textual or phonemic conditions.
\newpage
\bibliographystyle{IEEEbib}
\bibliography{refs}

\end{document}